\documentclass[aip,pof,amsmath,amssymb,reprint,showpacs] {revtex4-1} 

\usepackage{graphicx}
\usepackage{dcolumn}
\usepackage{bm}
\usepackage{booktabs}

\newcommand{\diff}[2]{\frac{\partial #1}{\partial #2}}

\newcommand{\klamm}[1]{\left(#1\right)}
\newcommand{\klammCurl}[1]{\left\{ #1 \right\}}
\newcommand{\klammSquare}[1]{\left[ #1 \right]}

\newcommand{\defi}{{\mathrel{\mathop:}=}}

\newcommand{\nDensity}{n}

\newcommand{\chemPot}{\mu}

\newcommand{\depthLJ}{\varepsilon}
\newcommand{\depthLJW}{\depthLJ_{\text{w}}}
\newcommand{\LJdiam}{\sigma}

\newcommand{\surfaceTension}{\gamma}

\newcommand{\pos}{{\bf r}}

\newcommand{\DisjoiningPressure}{\Pi}
\newcommand{\chemPotN}{\chemPot}

\newcommand{\nDensityV}{\nDensity_{\text{vap}}}
\newcommand{\nDensityL}{\nDensity_{\text{liq}}}
\newcommand{\nDensityW}{\nDensity_{\text{w}}}

\newcommand{\surfaceTensionLV}{\surfaceTension_{\text{lv}}}
\newcommand{\surfaceTensionWV}{\surfaceTension_{\text{wv}}}
\newcommand{\surfaceTensionWL}{\surfaceTension_{\text{wl}}}

\newcommand{\filmThickness}{\ell}
\newcommand{\GrandPotential}{\Omega}

\newcommand{\FE}{\mathcal{F}}
\newcommand{\FEhs}{\mathcal{F}_{\text{HS}}}
\newcommand{\FEattr}{\mathcal{F}_{\text{attr}}}

\newcommand{\Vext}{V_{\text{ext}}}

\newcommand{\BHattr}{\phi_{\text{attr}}}

\newcommand{\BHattrWF}{\phi_{\text{attr}}^{\text{wf}} }

\newcommand{\dI}{\text{d}}

\newcommand{\new}[1]{#1}

\begin{document}

\preprint{AIP/123-QED}

\title{Fluid structure in the immediate vicinity of an equilibrium three-phase contact line and assessment of disjoining pressure models
using density functional theory}

\author{Andreas Nold}
\email{andreas.nold09@imperial.ac.uk}
\affiliation{Department of Chemical Engineering, Imperial College London, London SW7 2AZ, United Kingdom}

\author{David N. Sibley}
\email{d.sibley@imperial.ac.uk}
\affiliation{Department of Chemical Engineering, Imperial College London, London SW7 2AZ, United Kingdom}

\author{Benjamin D. Goddard}
\email{b.goddard@ed.ac.uk}
\affiliation{The School of Mathematics and Maxwell Institute for Mathematical Sciences, The University of Edinburgh, Edinburgh EH9 3JZ, United Kingdom}

\author{Serafim Kalliadasis}
\email{s.kalliadasis@imperial.ac.uk}
\affiliation{Department of Chemical Engineering, Imperial College London, London SW7 2AZ, United Kingdom}

\date{\today}

\begin{abstract}

We examine the nanoscale behavior of an equilibrium three-phase contact line
\new{in the presence of long-ranged intermolecular forces}
by employing a statistical mechanics of fluids approach, namely density
functional theory (DFT) together with fundamental measure theory (FMT). This
enables us to evaluate the predictive quality of effective Hamiltonian models
in the vicinity of the contact line. In particular, we compare the results
for mean field effective Hamiltonians with disjoining pressures defined
through (I) the adsorption isotherm for a planar liquid film, and (II) the
normal force balance at the contact line. We find that the height profile
obtained using (I) shows good agreement with the adsorption film thickness of
the DFT-FMT equilibrium density profile in terms of maximal curvature and the
behavior at large film heights. In contrast, we observe that while the
height profile obtained by using (II) satisfies basic sum rules, it shows
little agreement with the adsorption film thickness of the DFT results. The
results are verified for contact angles of $20^\circ$, $40^\circ$ and
$60^\circ$.
\end{abstract}

\maketitle

\section{Introduction}

Consider a droplet sitting on a substrate and surrounded by its saturated
vapor.  In the partial wetting regime, the droplet's surface meets the
substrate at a finite contact angle. Macroscopically, this contact angle is a
material constant of the fluid-wall pair. Microscopically, intermolecular
forces dictate the exact structure of the fluid in the vicinity of the point
where the liquid, the vapor phase and the substrate meet. We study this
microscopic structure using elements from the statistical mechanics of fluids
and compare our results with two approaches \new{using}
coarse-grained mean-field Hamiltonian theory.

Understanding the exact structure at a contact line with a finite contact
angle is of increasing interest for a wide spectrum of technological
applications but also from a fundamental point of view. Recent advances in technology allow the design of devices of increasingly small size, highlighting the importance of developing a
fundamental understanding of phenomena at small scales for the manipulation and control of fluids in micro-/nanofluidic devices.\cite{Seemann08022005,Rauscher2008} Small-scale phenomena in wetting
are also important in biology, e.g. the rupture of liquid films in the
airways of lungs,\cite{jensen1992insoluble} or the rupture of the tearfilm on
the eyeball.\cite{Wong199644} Finally, a well-founded understanding of
equilibrium behavior is a prerequisite for the accurate modeling of the
dynamic contact line behavior.

The study of the microscopic structure at the contact line is limited by its
high computational cost.\cite{macdowell2011computer} Two ways of computing
the density structure for equilibrium systems are Monte-Carlo (MC)
\cite{Herring:2010vn,Das:2010:MonteCarloNanoscale} and Molecular Dynamics
(MD) computations.\cite{Werder:2001,Tretyakov2013parameter} MC and MD solve
for the positions of individual particles, such that the number of particles
necessarily limits the system size to nanoscales, and despite dramatic
improvements in computational power, \new{MC/MD} computations are still only
applicable for small fluid volumes. As an alternative to particle-based
computations, classical density functional theory (DFT) allows to solve
directly for the density distribution of inhomogeneous systems
\cite{Evans,Wu-DFT} and retains the microscopic details of macroscopic
systems but at a cost much lower to that in \new{MC/MD}. In the past, DFT has been
predominantly applied to one-dimensional (1D) scenarios
\cite{Stewart:2005tg,Nold:2011zr} but has recently been used in
two-dimensional (2D) scenarios such as nanodrops,
\cite{RuckensteinNanorough,NanodropPlanarSurface} critical point wedge filling,~\cite{Malijevsky:2013:CriticalPointWedgeFilling} capillary
prewetting,~\cite{PeterPRE} and three-dimensional nucleation processes.~\cite{Zhou:2012}

Intermolecular interactions between polar and non-polar particles, as well as
hydrogen bonds and other interactions, are short-ranged and decay
exponentially.\cite{Bonn.20090527} Physically, however, the much more common
occurrence are fluids with apolar, uncharged particles with long-range van
der Waals type interactions \cite{Beguin:2013} --- which appear to be the
`only general aspect of the physics of wetting that is not well-understood'.
\cite{Henderson:2005dz} Fluids with long-range interactions exhibit a
different wetting behavior to fluids with short-ranged interactions.
\cite{Bonn.20090527} In addition, the latter are numerically more
accessible as they allow for a finite cutoff length for the inter-particle
interactions.

Wetting of fluids with long-range dispersion forces has been studied by means
of a sharp-kink approximation for the density
profile.\cite{Dietrich,Bauer:1999ys,Hofmann:2010:Nanopatterned} Other studies
have relaxed the density profile by using a local density approximation for
the hard-sphere inter-particle potential.\cite{Antonio2010} However, at
present the most successful and accurate DFT for hard-sphere systems with
attractive interactions is that of fundamental measure theory
(FMT).\cite{Roth:2010fk} DFT-FMT with dispersion forces has been successfully
applied \new{in studies} of critical point wedge
filling~\cite{Malijevsky:2013:CriticalPointWedgeFilling}
\new{and density computations in the vicinity of liquid wedges.~\cite{Merath:2008}}

Here we construct an equilibrium three-phase contact line of a fluid with
dispersion interactions in the immediate vicinity of a wall using an accurate
FMT model for the hard-sphere interactions. This allows us to solve for the
density profile of the fluid in the contact line region and to shed light on
the fluid structure there down to the nanoscale: In particular, we observe
the fluid structure there down to the nanoscale: in particular, we observe
the presence of a step-like structure for the higher fluid densities very
close to the contact line.
The DFT-FMT approach also allows us to test mean-field effective Hamiltonian
approaches which reduce the dimension of the system by one, describing the
interface by a simple height profile. In general, the Hamiltonian of such
systems is written as a sum of the contribution due to the liquid-vapor
interface and an effective interface
potential.\cite{Lipowsky:1987,Mikheev:1991pi} 

The interface potential goes back to the concept of disjoining pressure
introduced by Derjaguin
\cite{Derjaguin:1987:50YearsOfSurfaFceScience,derjaguin1986properties} and
Frumkin \cite{Frumkin1938I} (but we note that Derjaguin's work was published
earlier~\cite{Derjaguin:1936}). The disjoining pressure is defined as the
excess pressure acting on a substrate due to the presence of a thin liquid
film, and it was directly linked with the adsorption isotherm, the plot of
the film thickness against the chemical potential of the system. Later,
Dzyaloshinskii, Lifshitz and Pitaevskii (DLP) directly computed the
disjoining pressure from the dispersion
interactions.\cite{Dzyaloshinskii:1961}
This connection between the definitions of disjoining pressure, as well as
its applicability to nonplanar systems in the framework of an effective
Hamiltonian theory, was recently analyzed in a number of discussion papers
that appeared in~{\it Eur. Phys. J.: Spec. Top.}, special issue `Wetting and
Spreading Science - quo
vadis?'.\cite{macdowell2011computer,Henderson:2011:EPJST:DisjoiningPressure,MacDowell:2011:ResponseEPJST,Henderson:EPJST:ResponseMacDowell,Henderson:NoteContinuingContactLine}
In particular, the transferability and/or universality, of results using
disjoining pressure in an effective Hamiltonian approach, as well as the
validity of the different definitions of the disjoining pressure were
questioned in these papers.

In this work, we make progress towards addressing these questions by
employing a DFT-FMT framework for a system with long-range wall-fluid and
fluid-fluid interactions. In particular, we study and directly compare two
routes to the disjoining pressure. First, we compute the adsorption isotherm
employing DFT in a planar configuration. Using an effective Hamiltonian
approach, this allows us to define a specific height profile across the
contact line. Secondly, we compute the full density distribution of a
three-phase contact line. This exact result can be used to define a
disjoining pressure based on the normal force balance,\cite{Herring:2010vn}
in the spirit of a parameter-passing technique. The disjoining pressure
containing this information from the 2D density profile is in turn inserted
into a Hamiltonian approach to compute a simple height profile.

In Sec. II, we give an overview of the DFT model we employ to solve for
the exact density profile in the vicinity of an equilibrium contact line. In
Sec. III, we give details of the numerical scheme we developed to solve the
DFT equations. A brief introduction to Hamiltonian approaches together with
the two definitions of the disjoining pressure considered in this study is
given in Sec. IV. In Sec. V we compare the DFT results with the Hamiltonian
approaches. Finally, we summarize our results and provide an outlook to
future work in Sec. VI.

\section{DFT Model}
\begin{figure}
	\centering	
	\includegraphics[width=8cm]{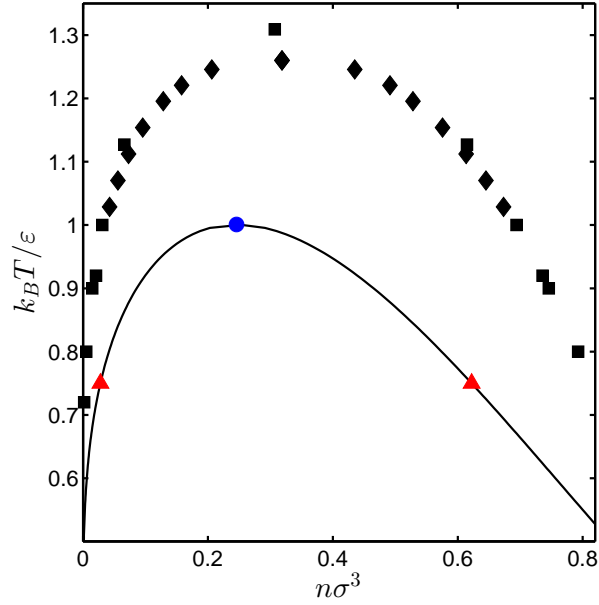}
		\caption{\new{Bulk phase diagram for values at saturation. The solid line represents the phase diagram for the model used in this work.
        The circle denotes the critical point at $\{k_B T_c = 1.0 \depthLJ, n_c
        \LJdiam^3 = 0.246\}$; the triangles denote the vapor and liquid densities
        for the temperature of $T = 0.75 T_c$ at which all computations in this work
        are done. The black squares denote experimental results for argon ($\LJdiam =
        3.405 \times 10^{-8}\text{cm}$ and $\depthLJ = 165.3 \times 10^{-16}
        \text{erg}$) by Michels, Levelt, and De Graaff.~\cite{Michels1958659} The
        black diamonds represent canonical MD simulations by
        Trokhymchuk and Alejandre.~\cite{Trokhymchuk:1999ye}}}
	\label{fig:BulkPhaseDiagram}
\end{figure}

We employ classical DFT to study the density distribution in the vicinity of
a static contact line. Classical DFT has been of paramount importance for the
study of inhomogeneous fluids. It is based on Mermin's theorem,
\cite{Mermin:1965lo} which allows the Helmholtz free energy $\FE$ to be
written as a unique functional of the number density profile $\nDensity({\bf
r})$.\cite{Wu-DFT} It can be shown rigorously that the equilibrium density
distribution minimizes the grand potential \cite{Evans}
\begin{align}
\new{
\GrandPotential[\nDensity] = \FE[\nDensity] + \int \nDensity(\pos) \klammCurl{\Vext({\bf r}) - \chemPotN} \dI\pos, } \label{eq:GrandPotential}
\end{align}
where $\chemPotN$ is the chemical potential and $\Vext$ is the external
potential. We minimize Eq.~(\ref{eq:GrandPotential}) by solving the
Euler-Lagrange equation
\begin{align}
\frac{\delta \Omega[\nDensity]}{\delta \nDensity({\pos})} = 0.\label{eq:EulerLagrangeEquation}
\end{align}
For a simple fluid of particles interacting with a Lennard-Jones (LJ) potential, the free energy is usually split into a repulsive hard-sphere part and an attractive contribution
\begin{align}
\FE[\nDensity] = \FEhs[\nDensity] + \FEattr[\nDensity].
\end{align}
We model the hard-sphere contribution with a Rosenfeld FMT approach,\cite{Rosenfeld:1989qc} which accurately models both structure and thermodynamics of hard-sphere fluids.\cite{Roth:2010fk}  The attractive interactions are modeled with a mean-field Barker-Henderson approach \cite{Barker:1967rq}
\begin{align}
\FEattr[\nDensity] &= \frac{1}{2 } \iint  \phi_{\text{attr}}({|\pos - \pos'|}) \nDensity(\pos)\nDensity(\pos') \dI\pos' \dI\pos, \label{eq:FEattr}
\end{align}
where the attractive interaction potential is given by
\begin{align}
\BHattr\klamm{r} =
\depthLJ \left\{ \begin{array}{ll}
0 & \text{for } r \leq  \LJdiam\\
4 \klamm{
\klamm{\frac{\LJdiam}{r}}^{12} -
\klamm{\frac{\LJdiam}{r}}^{6}
} & \text{for } r > \LJdiam
\end{array}  \right. .
\end{align}
Here, $\LJdiam$ is the distance from the center of the particle at which the
LJ potential is zero and $\depthLJ$ is the depth of the LJ potential. The
simple fluid described by the given model has a critical point at $k_B T_c =
1.0 \depthLJ$, where $k_B$ is the Boltzmann constant and all computations in
this work were performed at $T = 0.75 T_c$, at which the liquid and vapor
number densities are well-separated ($\nDensityL \LJdiam^3= 0.622$,
$\nDensityV \LJdiam^3= 0.003$) \new{and at which the surface tension becomes
$\surfaceTensionLV = 0.3463 \depthLJ/\LJdiam^2$}. All 2D computations are
performed at the saturation chemical potential, at which the bulk vapor and
bulk liquid are equally stable. \new{A phase diagram of the model used in
this work is depicted in Fig.~\ref{fig:BulkPhaseDiagram}, and compared with
experimental and simulation results for argon.~\cite{Michels1958659,Trokhymchuk:1999ye} We note that the discrepancy
between the data stems from the fact that argon is not well modeled with a
Barker-Henderson interaction potential. For a DFT model which reproduces more accurately
the bulk properties of argon, see e.g. Peng and Yu.~\cite{Peng:2008}}

The external potential is derived from the interaction of the
wall-fluid particles, modeled analogously to the fluid-fluid interaction as
\begin{align}
\BHattrWF\klamm{r} =
\depthLJW \left\{ \begin{array}{ll}
\infty & \text{for } r \leq  \LJdiam\\
4 \klamm{
\klamm{\frac{\LJdiam}{r}}^{12} -
\klamm{\frac{\LJdiam}{r}}^{6}
} & \text{for } r > \LJdiam
\end{array}  \right. , \label{eq:WallFluidInteraction}
\end{align}
where $\depthLJW$ is the depth of the wall-fluid interactions. Consider a
Cartesian coordinate system with the $x$-$z$ plane parallel to the wall
and  the $y$-coordinate direction normal to the wall. The external potential is
then obtained from the integration of the interactions over the uniform
density distribution of wall particles $\nDensityW$ for $y \leq -\LJdiam$,
giving
\begin{align}
\Vext\klamm{y}
&=
\left\{
\begin{array}{ll}
\infty & y \leq 0\\
\frac{2}{3}\pi \alpha_{\text{w}} \LJdiam^3 \klammSquare{\frac{2}{15} \klamm{\frac{\LJdiam}{y + \LJdiam} }^9 - \klamm{\frac{\LJdiam }{y+ \LJdiam}}^3} & y > 0
\end{array}
\right., 
\end{align}
where $\alpha_{\text{w}} = \nDensityW \depthLJW$ is the strength of the wall potential.

\section{Computations}

\begin{figure}
	\centering		
		\includegraphics[width=8cm]{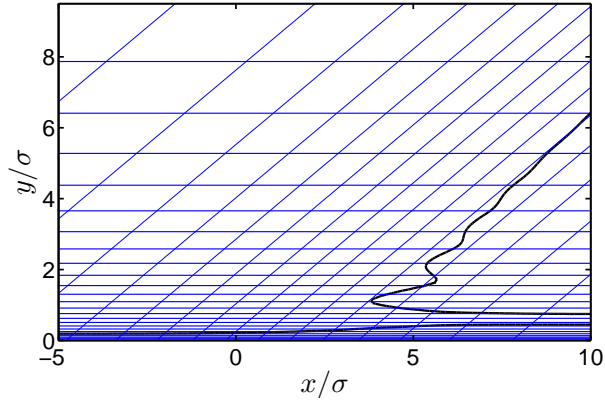}
		\caption{\new{Sketch of part of the grid employed in the numerical computations of this work.
        The complete grid covers the full half space $y>0$. The gridlines represent isolines of the computational
        variables $\xi$ and $\eta$, which are mapped to the physical space $(x,y)$ as given in Eqs.~(\ref{eq:MappingCompToXPrime})
        and (\ref{eq:MappingXPrimeToX}). Here we plot every second isoline for a grid with
        $45$ and $75$ Chebychev collocation points in the $\xi$ and $\eta$ direction, respectively,
        and with parameters $L_1 = 4 \LJdiam$, $L_2 = 2\LJdiam$ and angle $\theta_{\text{n}} = 40^\circ$.
        The black solid line represents one isoline of a density profile for a contact line computation.}}
	\label{fig:ExemplaricGrid}
\end{figure}

\new{Solving for the full microscopic density profile at the contact line requires
a considerable amount of modeling and computational effort, restricting
computations to systems of very small size, such as
nano-droplets.\cite{RuckensteinNanorough,NanodropPlanarSurface} In the
configuration we discuss here, this is circumvented by constructing a liquid
wedge (at saturation) in contact with the substrate and with a well-defined
three-phase contact line. This effectively allows us to model the contact
line of a macroscopic droplet.}

In this case, choosing a skewed grid for a representation of our numerical results,
such as depicted in Fig.~\ref{fig:ExemplaricGrid}, is computationally
advantageous. For a map from the computational to the physical domain, we
employ a spectral collocation method \cite{Trefethen_2000} to represent
functions in the half space $y \geq 0$. In particular, we employ a tensor
product of two 1D Chebychev grids in $\klamm{\xi,\eta} \in
[-1,1]\times[-1,1]$. This domain is mapped onto the half-space through
\cite{Shen2009}
\begin{align}
 x' = L_1 \frac{\xi}{\sqrt{1 - \xi^2}} ,\qquad
 y' = L_2 \frac{1+\eta}{1-\eta}, \label{eq:MappingCompToXPrime}
\end{align}
where $L_1,L_2$ represent the length-scales of the map. 
The maps are such that half of the collocation points in each direction $\xi$, $\eta$ are mapped onto the intervals $[-L_1,L_1]$ and $[0,L_2]$, respectively.
In order to efficiently represent density distributions of wedges with relatively small contact angles, the grid is skewed by an angle $\theta_{\text{n}}$, such that
\begin{align}
 x = \frac{x'}{\sin \theta_{\text{n}}} + y' \cot\theta_{\text{n}},\qquad
 y = y', \label{eq:MappingXPrimeToX}
\end{align}
where division by $\sin \theta_{\text{n}}$ corrects the scaling of $L_1$ in
the skewed grid, such that the number of collocation points across a
liquid-vapor interface at angle $\theta_{\text{n}}$ is invariant with
respect to the angle. We then impose that the density at the collocation
points for $y > y_{\text{max}}$ corresponds to a straight wedge with angle
$\theta_{\text{n}}$. In other words, the angle of the liquid-vapor interface
for $y > y_{\text{max}}$ is imposed as a boundary condition.
\new{Also, we consider density profiles which converge smoothly to the planar wall-vapor and wall-liquid equilibrium
density profiles as $x \to - \infty$ and $x \to \infty$, respectively.}

\begin{figure}
	\centering		
		\includegraphics[width=8cm]{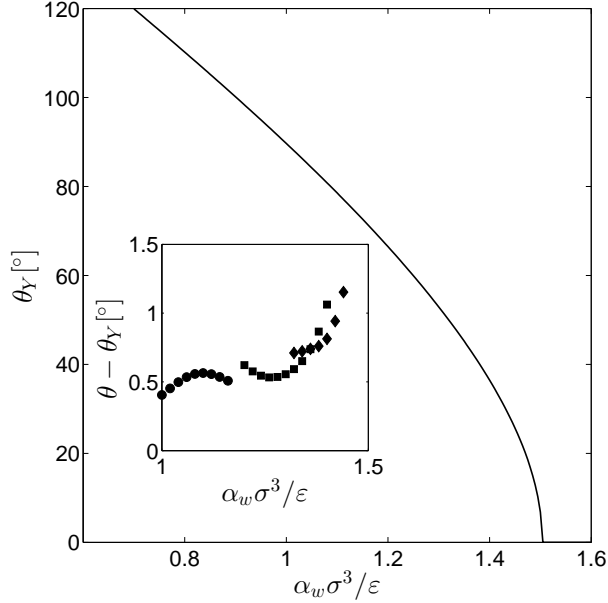}
	\caption{Plot of the contact angle dependence on the strength of the wall attraction $\alpha_{\text{w}}$, computed via Young's equation (\ref{eq:YoungsEquation}) \new{from surface tensions as calculated from DFT for planar geometries}. Complete wetting is reached at $\alpha_{\text{w}}\LJdiam^3/\depthLJ = 1.50$.
	The inset shows measurements of the contact angle from 2D computations as described in the text with $y_{\text{max}} = 15\LJdiam$ as deviations from the Young contact angle. Circles, squares and diamonds depict computations with $\theta_{\text{n}}=90^\circ$, $60^\circ$ and $40^\circ$, respectively.}
	\label{fig:Sketch_WettingEPW_THETA}
\end{figure}

Physically, the contact angle \new{$\theta_{\text{Y}}$} of a liquid-vapor interface in contact with a substrate is uniquely defined by the fluid properties and the external potential induced by the substrate through Young's equation
\begin{align}
\surfaceTensionLV \cos \theta_{\text{Y}} = \surfaceTensionWV - \surfaceTensionWL, \label{eq:YoungsEquation}
\end{align}
where $\surfaceTensionLV$ is the liquid-vapor surface tension and $\surfaceTensionWL$ and $\surfaceTensionWV$ are the wall-liquid and the wall-vapor surface tensions. In Fig.~\ref{fig:Sketch_WettingEPW_THETA}, we plot the Young contact angle \new{as a function of the wall attraction $\alpha_{\text{w}}$,} where the surface tensions $\surfaceTensionLV$, $\surfaceTensionWL$ and $\surfaceTensionWV$ were obtained from planar DFT computations.

\begin{figure}
	\centering		
		\includegraphics[width=8cm]{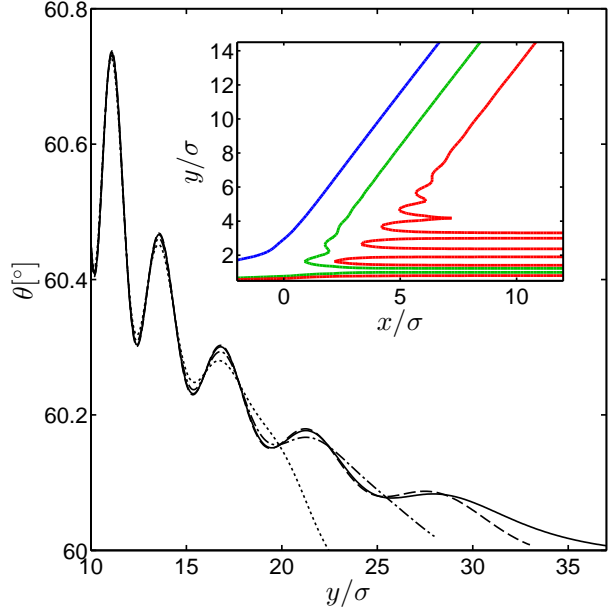}
		\caption{Slope of the isodensity line for
        $\nDensity = \klamm{\nDensityV + \nDensityL}/2$ for $y_{\text{max}} = \{ 20,25,30,35\}$, represented by the dotted,
        dash-dotted, dashed and solid lines, respectively. Computations are done on a
        grid with $\theta_{\text{n}} = 60^\circ$. The substrate strength is
        $\alpha_{\text{w}} \sigma^3/\depthLJ = 1.25$, such that $\theta_{Y}
        = 60.0^\circ$. The inset depicts a contour plot of the contact line region. The contour lines correspond to number densities
        $(\nDensity-\nDensityV)/(\nDensityL-\nDensityV) =\{0.05,0.5,0.95\}$ from left to right, respectively.}
	\label{fig:y2MaxAsymptotics60}		
\end{figure}

The Young contact angle based on planar DFT computations is then compared
with measurements of the contact angle in 2D settings. As described above,
for $y > y_{\text{max}}$, the contact angle is fixed numerically to
$\theta_{n}$, while for $y < y_{\text{max}}$, the contact angle formed by the
liquid-vapor interface at equilibrium should correspond to the Young contact
angle. To test this and ensure that the values measured do not depend on the
numerical parameters $\theta_n$ and $y_{\text{max}}$, we measure the contact
angle in two steps. First, we compute equilibrium configurations for
orthogonal ($\theta_{\text{n}}$ equal $90^\circ$) and skewed grids with
$\theta_{\text{n}}$ equal $40^\circ$ and $60^\circ$ and for $y_{\text{max}} =
15\LJdiam$. Measuring the average slope of the isodensity line for $\nDensity = \klamm{\nDensityV +
\nDensityL}/2$ in the interval $y \in [10\LJdiam,14\LJdiam]$ then allows us
to get a rough first estimate for the physical contact angle. In the cases
considered here, the slope of the height profile asymptotically approaches
the slope dictated by the Young contact contact angle from above, which means
that the measured average slope leads to an overestimation of the contact
angle (see inset of Fig.~\ref{fig:Sketch_WettingEPW_THETA}). In a next step,
we choose three specific substrate strengths, and set $\theta_{\text{n}}$
equal to the estimated contact angle. We then increase $y_{\max}$ and check
if this affects the numerical results for the slope of the height profile in
the vicinity of the contact line. In an iterative procedure,
$\theta_{\text{n}}$ is adjusted such that no dependency on $y_{\max}$ is
observed. An example study of the slope dependence on the numerical parameter
$y_{\max}$ is depicted in Fig.~\ref{fig:y2MaxAsymptotics60}. This procedure
leads to a set of numerical parameters which allows for a very efficient
representation and analysis of density distributions of contact lines in a
wide range of contact angles.

\new{We note that to solve the Euler-Lagrange Eq.~(\ref{eq:EulerLagrangeEquation}), it is necessary to
compute the functional derivative of $\FEattr[\nDensity]$. This corresponds to a convolution of the
density profile $\nDensity(\pos)$ with the attractive interaction potential $\BHattr\klamm{|\pos|}$, given by
\begin{align}
\frac{\delta \FEattr[\nDensity]}{\delta \nDensity \klamm{\pos }} &= \int  \BHattr({|\pos - \pos'|}) \nDensity(\pos') \dI\pos'. \label{eq:Convolution}
\end{align}
Let us describe briefly how this expression is computed numerically at a
collocation point $\pos$. It is worth noting that $\BHattr(r)$ vanishes for
$r \leq \sigma$. Hence, for each point $\pos = (x,y)$, an area given by
$\mathcal{A}_\pos = \{\pos' = (x',y') \in \mathbb{R}^2| y'>0, |\pos' - \pos| >
\sigma \}$ is discretized. Depending on the value of $y$, $\mathcal{A}_\pos$
is divided into two or three subareas which are discretized separately using
spectral collocation methods. We emphasize that technically, by employing a
spectral method and placing collocations on the full area $\mathcal{A}_\pos$,
we do not introduce a cutoff for $\BHattr$. This is particularly convenient
for our choice of long-range fluid-fluid interactions. The numerical accuracy
is instead limited by the quality of the maps used to discretize
$\mathcal{A}_\pos$ - including the choice of mapping parameters, number of
collocation points, and the quality of the discretization of
$\nDensity(\pos)$ on the original grid. After interpolating the values of the
density onto the collocation points of $\mathcal{A}_\pos$, the density is
multiplied with $\BHattr$ such as given in (\ref{eq:Convolution}) and the
integration is performed on $\mathcal{A}_\pos$. The procedure is repeated for
each collocation point. The result of the interpolation and subsequent
multiplication and integration is assembled in a convolution matrix in a
preprocessing step.}

All computations were performed using Matlab on a Intel Core i7-3770, 3.4GHz
desktop PC with 8GB RAM running Windows 7. Preprocessing the FMT integration
matrices and the convolution matrices for a grid with a specific
$\theta_{\text{n}}$ with $50\times 80$ collocation points takes approximately
3.5h. Solving the Euler-Lagrange Eq.~(\ref{eq:EulerLagrangeEquation})
for a specific $y_{\max}$ takes 0.5h--2h depending on the specific
configuration.

\section{Hamiltonian approaches and disjoining pressure}

Computations for full macroscopic systems such as macroscopic droplets
require a coarse-grained approach. One way to retain essential information of
the structure of the contact line without computing the full density
profile is through interface Hamiltonian approaches which reduce the
dimension by one.\cite{Lipowsky:1987,Mikheev:1991pi} In particular, for
systems which are not too close to the critical point, the film height
profile of the liquid-vapor interface $h(x)$ can be studied by minimizing
the Hamiltonian \cite{Herring:2010vn}
\begin{align}
H[h] = \int_{-\infty}^\infty \klammCurl{ \surfaceTensionLV \klamm{ \sqrt{ 1+ (h')^2 } -1 } + V(h)  } \dI x, \label{eq:Hamiltonian_h}
\end{align}
where $h' = {\dI h}/{\dI x}$ is the slope of the interface and $V(h)$ is the effective interface potential. Minimising the Hamiltonian with respect to $h(x)$ leads to the defining equation for the height profile
\begin{align}
-\DisjoiningPressure(h(x)) =  \surfaceTensionLV \frac{\dI}{\dI x} \klamm{ \frac{h'(x)}{\sqrt{1+\klamm{h'(x)}^2}}}, \label{eq:FilmHeightAndDisjoiningPressure}
\end{align}
where the disjoining pressure is the negative derivative of the interface potential
\begin{align}
\DisjoiningPressure\klamm{h}  \defi - \frac{\dI V}{\dI h}. \label{eq:DisjoiningPressureVh}
\end{align}
Integrating Eq.~(\ref{eq:FilmHeightAndDisjoiningPressure}) along the
coordinate parallel to the wall leads to
\begin{align}
-\int_{-\infty}^x \DisjoiningPressure(\hat x) \dI\hat x &=  \surfaceTensionLV \klamm{ \frac{h'}{\sqrt{1+(h')^2}}} \notag\\
&= \surfaceTensionLV  \sin \theta(x), \label{eq:DisjoiningPressure_IntX}
\end{align}
where it was used that the film converges to a \new{constant} wall-vapor film with film
height $\lim_{x\to-\infty}h(x) = h_0$. For $x\to \infty$, this corresponds
with the sum rule representing the normal force balance from Young's equation
\cite{derjaguin1986properties}
\begin{align}
-\int_{-\infty}^\infty \DisjoiningPressure(h(x)) \dI x = \surfaceTensionLV \sin \theta_{\text{Y}}, \label{eq:SumRuleNormalForceBalance}
\end{align}
where it was used that the film converges to a wedge with the Young contact
angle $\lim_{x \to \infty} h'(x) = \tan \theta_{\text{Y}}$. Similarly,
integrating (\ref{eq:FilmHeightAndDisjoiningPressure}) with respect to the
film height $h$ gives
\begin{align}
\int_{h_0}^{h(x)} \DisjoiningPressure(\hat h) \dI \hat h
&=\surfaceTensionLV
\left[
\frac{1}{\sqrt{1+\klamm{h'(\hat x)}^2}}
\right]_{\hat x=-\infty}^{\hat x = x} \notag\\
&= \surfaceTensionLV \klamm{\cos \theta(x) -1}. \label{eq:IntegratedFilmHeightDisjoiningPressure}
\end{align}
Integrating up to $h=\infty$ yields the important expression from Derjaguin-Frumkin theory \cite{derjaguin1986properties}
\begin{align}
- \int_{h_0}^\infty \DisjoiningPressure\klamm{h} \dI h = \surfaceTensionLV \klamm{1 - \cos \theta_{\text{Y}}}. \label{eq:FrumkinResultDisjoiningPressure1}
\end{align}
We note that the sum rules (\ref{eq:SumRuleNormalForceBalance}) and (\ref{eq:FrumkinResultDisjoiningPressure1}) hold independently of the exact definition of the interface potential.

\new{An accurate model for the interface potential $V$ is crucial
in order to retain important information about the structure of the contact
line. Usually, $V(h)$ is defined as the interface potential for a planar film
of height $h$.\cite{Bonn.20090527} In the last decade, Hamiltonian models
have been suggested which take into account nonlocal effects due to changes
of the height profile along the substrate. These include nonlocal models for
short-ranged wetting,\cite{parry2006derivation} as well as models which
include the slope of the height profile in the interface
potential.\cite{Dai:2008} We now compare one local model of the disjoining
pressure with a model based on a full DFT computation of a liquid wedge and
test their predictive capabilities.}

\subsection{Adsorption isotherm}
For a planar liquid film and under the assumption that the free energy is a
function of the film thickness
only,\cite{derjaguin1986properties,Frumkin1938I} the grand potential per unit
area can be reduced to
\begin{align}
\surfaceTension\klamm{\filmThickness,\chemPotN} = f(\filmThickness) - \chemPotN \filmThickness \Delta \nDensity, \label{eq:ReducedFreeEnergy}
\end{align}
where $\Delta \nDensity = \nDensityL - \nDensityV$.
\new{We note that Eq.~(\ref{eq:ReducedFreeEnergy}) can be derived from the general formulation of the grand potential (\ref{eq:GrandPotential}) by assuming a dependence of
the density profile $\nDensity(y)$ on the film thickness $\filmThickness$ but
without any dependence on the chemical potential. One method to do this is
through a simple sharp-interface approximation. $f(\ell)$ is then the reduced
form of the part $\FE[\nDensity] + \int \nDensity(\pos) \Vext({\bf r})
\dI\pos$ in Eq.~(\ref{eq:GrandPotential}).} At equilibrium, $\filmThickness$
minimizes $\surfaceTension$. Let us define
$\chemPotN_{\text{eq}}(\filmThickness)$ as the chemical potential at which a
film of thickness $\filmThickness$ is at equilibrium
\begin{align}
\chemPotN_{\text{eq}}(\filmThickness)  \defi \frac{1}{\Delta \nDensity} \new{\frac{\dI f}{\dI\filmThickness}}.
\end{align}
In the planar case, $\surfaceTension$ corresponds to the effective interface potential $V(h)$, and following Frumkin's derivation,\cite{Frumkin1938I}
the disjoining pressure is defined by the negative derivative of this quantity with respect to film thickness, leading to
\begin{align}
\DisjoiningPressure_{\text{I}}\klamm{\filmThickness,\chemPotN} \defi - \diff{\surfaceTension}{\filmThickness}
= \klamm{\chemPotN - \chemPotN_{\text{eq}}\klamm{\filmThickness}} \Delta \nDensity, \label{eq:DisjoiningPressureAdsorptionIsotherm}
\end{align}
linking the disjoining pressure with the adsorption isotherm.\cite{Henderson:2005dz} We note that by definition the disjoining pressure
is zero for the equilibrium film thickness, \new{consistent with $\chemPotN$}. In other words, the disjoining
pressure gives the excess pressure acting on the substrate for liquid films
which are perturbed or off-equilibrium, e.g.~through fluctuations or because
of forcing through boundary conditions.

\begin{figure}
	\centering		
		\includegraphics[width=8cm]{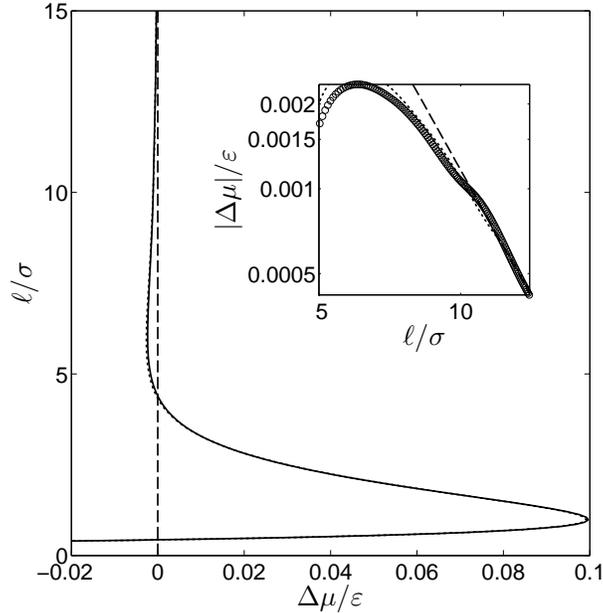}
	\caption{Adsorption isotherm for $\alpha_{\text{w}} \LJdiam^3/\depthLJ = 1.375$ ($\theta_{\text{Y}}= 41.1^\circ$). The solid line represents the film thickness
    (\ref{eq:AdsorptionFilmThickness}) of equilibrium density distributions, and the dotted line represents \new{the modified} sum rule \new{(\ref{eq:SumRuleDisjoiningPressue:Mod})}. The contact angle obtained from sum rule (\ref{eq:FrumkinResultDisjoiningPressure1}) is $41.1^\circ$, in excellent agreement with the Young contact angle $\theta_{\text{Y}}$. The inset shows the asymptotic behavior
	for $\filmThickness\to\infty$, as $\Delta \mu \sim \filmThickness^{-3}$. The dashed line is a fit
    for $\filmThickness \in [10 \LJdiam,15 \LJdiam]$, giving $\Delta \mu = a
    \filmThickness^{-3}$ with the coefficient $a = -1.16 \depthLJ \LJdiam^3$. The
    circles in the inset represent the individual DFT computations of the
    equilibrium density, which in the main plot are connected by the solid line
    for convenience.}
	\label{fig:AdsorptionIsotherm41}
\end{figure}

\subsection{Normal force balance}
Dzyaloshinskii, Lifshitz and Pitaevskii (DLP) employed quantum-field theory to directly compute the force acting on the surface due to an adsorbed film, and related it to the disjoining pressure.\cite{Dzyaloshinskii:1961}
 In other words, DLP related the chemical potential difference (\ref{eq:DisjoiningPressureAdsorptionIsotherm}) to the excess pressure on the substrate wall due to an adsorbed liquid film. For a discussion of this connection, see also Ref.~\onlinecite{Henderson:2011:EPJST:DisjoiningPressure}. In particular, the force acting on the substrate at \new{saturation chemical potential $\chemPotN_{\text{sat}}$ for a density profile $\nDensity_\filmThickness$} is given by  \cite{Mikheev:1991pi,Henderson:2005dz}
\begin{align}
\new{\DisjoiningPressure(\filmThickness) = - \int_{-\infty}^\infty  \klamm{ \nDensity_\filmThickness(y) - \nDensity_{\filmThickness = \infty}(y) } V_{\text{ext}}'(y) \dI y,}\label{eq:SumRuleDisjoiningPressue}
\end{align}
where $\nDensity_\filmThickness(y)$ is \new{a} density profile
\new{at chemical potential $\chemPotN_{\text{sat}}$ but with the additional constraint of film thickness $\filmThickness$. Such profiles may be either partially stable or unstable, and are obtained by minimizing the 
excess grand potential subject to the constraint of fixed adsorption.~\cite{Henderson:2005dz}} $\nDensity_{\filmThickness = \infty}(y)$ is thus the density profile of the
equilibrium case of a film of infinite thickness accounting for the
contribution from the bulk pressure as \cite{Henderson1992}
\begin{align}
\new{p_{\text{sat}} = - \int_{-\infty}^\infty \nDensity_{\filmThickness=\infty}(y) V_{\text{ext}}'(y) \dI y}. \label{eq:SumRule:PressureDensityProfile}
\end{align}
\new{We note that in Eq.~(\ref{eq:SumRuleDisjoiningPressue}),
the disjoining pressure would decay exponentially if both the fluid-fluid and
the fluid-substrate interactions were short-range. Here, however, the
long-range fluid-substrate interactions lead to an algebraic decay of the
disjoining pressure. Furthermore, the interplay between long-range
fluid-fluid and the short-range part of the fluid-substrate interactions
also} lead to \new{an algebraic contribution to the disjoining pressure with
an} identical power series in terms of film
thickness $\filmThickness$.\cite{Henderson:2005dz} This means that one cannot
apply asymptotic theory to derive a distinct representation for long-range
and short-range interactions.\cite{Henderson:2005dz} While DLP circumvent
this problem by only applying their theory to films of mesoscopic
scales,\cite{Dzyaloshinskii:1961} we will instead use the full numerical
solution of the disjoining pressure in order to define a height profile
through the three-phase contact line.

\section{Results and discussion}

In Fig.~\ref{fig:AdsorptionIsotherm41}, we compare the Derjaguin-Frumkin
route of the disjoining pressure
(\ref{eq:DisjoiningPressureAdsorptionIsotherm}) and the definition from the
normal force balance for a planar film on a solid substrate. We have employed a numerical continuation scheme
to compute the full bifurcation diagram for the adsorption isotherm including
its meta- and unstable branches. As an order parameter for the number density
distribution, we have used the adsorption film thickness:
\begin{align}
\filmThickness \defi \frac{1}{\Delta \nDensity} \int_0^\infty \klamm{\nDensity(y) - \nDensityV} \dI y, \label{eq:AdsorptionFilmThickness}
\end{align}
where the vapor density $\nDensityV$ is taken at the chemical potential at which $n(y)$ is in equilibrium.
In the large film thickness limit, dispersion forces enforce an algebraic approach of the saturation line \cite{Dietrich} as
\begin{align}
\Delta \chemPotN  \sim \filmThickness^{-3} \qquad \text{for} \quad \filmThickness \to \infty,
\end{align}
where $\Delta \chemPotN = \chemPotN - \chemPotN_{\text{sat}}$ is the deviation of the chemical potential from its saturation value.
\new{Note that the density profiles obtained when solving for the adsorption isotherm are not computed at saturation chemical potential,
and can therefore not be used as $\nDensity_\filmThickness$ in Eq.~(\ref{eq:SumRuleDisjoiningPressue}). 
To allow for a comparison of the two routes to the disjoining pressure, 
we have instead combined Eqs.~(\ref{eq:SumRuleDisjoiningPressue}) and (\ref{eq:SumRule:PressureDensityProfile}) to define a generalized form of sum rule~(\ref{eq:SumRuleDisjoiningPressue}):
\begin{align}
\new{\DisjoiningPressure[\nDensity] = - \int_{-\infty}^\infty  \nDensity(y) V_{\text{ext}}'(y) \dI y - p_\infty,} \label{eq:SumRuleDisjoiningPressue:Mod}
\end{align}
where $p_\infty$ is the bulk pressure of the given density profile $\nDensity$ as $y \to \infty$.
The computations depicted in Fig.~\ref{fig:AdsorptionIsotherm41} give an excellent agreement between the two definitions.}

\begin{figure}[htbp]
\begin{center}
\includegraphics[width=8cm]{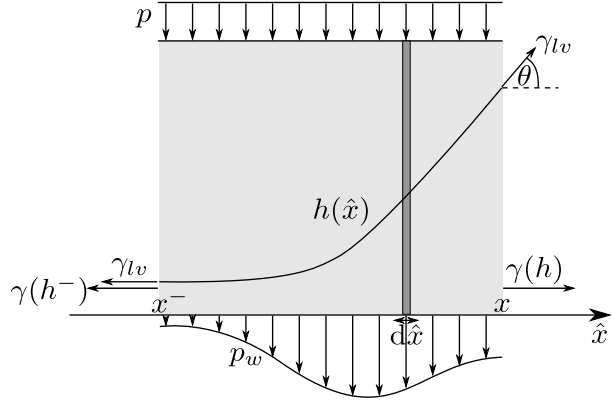}
\caption{Sketch of a mechanical model which describes the normal and the parallel force balances (\ref{eq:DisjoiningPressure_IntX}) and (\ref{eq:IntegratedFilmHeightDisjoiningPressure}) for an adsorbed liquid film in the box $[x^-,x]\times[0,\infty]$, where $x^-\to-\infty$. The force acting from the external potential modeling the substrate on the fluid strip $[\hat x,\hat x+\dI \hat x]\times[0,\infty]$ corresponds to the disjoining pressure $\DisjoiningPressure = p_w - p$ times the length of the interval $\dI \hat x$. Here, $p_w \dI\hat x = -\dI\hat x \int_{-\infty}^\infty  \nDensity(\hat x,y)  V_{\text{ext}}'(y) \dI y $ is the net force acting from the substrate on the fluid strip. The liquid-vapor surface tension $\surfaceTensionLV$ accounts for fluid-fluid interactions. 
The forces accounting for the fluid-fluid interactions stemming from the distortion of the density profile due to the substrate are included as a wall-liquid film-vapor potential $\surfaceTension(h)$, acting at the wall.}
\label{fig:SketchMechanicalModel}
\end{center}
\end{figure}

Let us now consider if there is a similar equivalence of disjoining pressure
definitions for the case of varying height profiles $h(x)$. For this purpose,
consider Eqs.~(\ref{eq:DisjoiningPressure_IntX}) and
(\ref{eq:IntegratedFilmHeightDisjoiningPressure}). We can formulate a
mechanical model for the contact line with height profile $h(x)$, in which
Eq.~(\ref{eq:DisjoiningPressure_IntX}) is a momentum balance in the direction
normal to the wall, and Eq.~(\ref{eq:IntegratedFilmHeightDisjoiningPressure})
is a momentum balance parallel to the wall for a system $(\hat x,\hat y) \in
[-\infty,x] \times [0,\infty]$ (see Fig.~\ref{fig:SketchMechanicalModel}).
Analogously, Eq.~(\ref{eq:FilmHeightAndDisjoiningPressure}) represents the
Young-Laplace equation modeling the pressure jump across a curved interface
in which the disjoining pressure acts as a gauge pressure in the liquid film.
The fact that the disjoining pressure in
Eq.~(\ref{eq:DisjoiningPressure_IntX}) represents the force of the substrate
acting on the fluid film allows the generalisation of
Eq.~(\ref{eq:SumRuleDisjoiningPressue}) to two dimensions by
\cite{Herring:2010vn,Henderson:2011:EPJST:DisjoiningPressure}
\begin{align}
\DisjoiningPressure_{\text{II}}\klamm{x} \defi - \int_{-\infty}^\infty  \klamm{ \nDensity(x,y) - \nDensity(\infty,y) } V_{\text{ext}}'(y) \dI y, \label{eq:SumRuleDisjoiningPressue2D}
\end{align}
where it was used that $\nDensity(x,y) V_{\text{ext}}'(y)$ is the force
acting through the external potential on the fluid element at point $(x,y)$,
and where the pressure acting from the bulk vapor was subtracted using
Eq.~(\ref{eq:SumRule:PressureDensityProfile}).

The definitions (\ref{eq:DisjoiningPressureAdsorptionIsotherm}) and
(\ref{eq:SumRuleDisjoiningPressue2D}) of disjoining pressures
$\DisjoiningPressure_{\text{I}}(h)$ and $\DisjoiningPressure_{\text{II}}(x)$,
respectively, allow in turn for the definition of two alternative height
profiles $h_{\text{I}}$ and $h_{\text{II}}$. Integrating
(\ref{eq:IntegratedFilmHeightDisjoiningPressure}) leads to the definition of
the film height profile through the ordinary differential equation
\begin{align}
h_{\text{I}}'  = \tan \klammCurl{{\cos^{-1}\klamm{1 + \frac{1}{\surfaceTensionLV}\int_{h_0}^{h_{\text{I}}} \DisjoiningPressure_{\text{I}}(\hat h)\dI\hat h}}}, \label{eq:2DDisjoiningPressureIntegrated_hP:H}
\end{align}
with boundary condition
\begin{align}
h_{\text{I}}\klamm{x_B} = h_B, \label{eq:BoundaryConditionHI}
\end{align}
for some $x_B,h_B$. Let us note that $h_{\text{I}}$ is translationally
invariant through the boundary condition. Integrating
Eq.~(\ref{eq:DisjoiningPressure_IntX}) leads to the height profile
\begin{align}
h_{\text{II}}'(x) =  -\tan\klammCurl{\sin^{-1}\klamm{\frac{1}{\surfaceTensionLV}\int_{-\infty}^{x} \DisjoiningPressure_{\text{II}}(\hat x)\dI\hat x}}, \label{eq:2DDisjoiningPressureIntegrated_hP}
\end{align}
with \new{boundary condition}
\begin{align}
\new{h_{\text{II}}(-\infty) = h_0,}
\end{align}
\new{ where $h_0$ is the (equilibrium) height of the vapor film. Note that while $h_{\text{I}}$ is translationally invariant through boundary condition (\ref{eq:BoundaryConditionHI}), $h_{\text{II}}$ is only invariant up to an additive constant, which does not change the position of the contact line in the direction parallel to the wall.}
Finally, we compare the film height profiles $h_{\text{I}}$ and $h_{\text{II}}$ with the adsorption film thickness
\begin{align}
h_{\text{III}}(x) = \frac{1}{\Delta \nDensity} \int_0^\infty \klamm{\nDensity(x,y) - \nDensityV} \dI y. \label{eq:Def_HIII}
\end{align}
	
\begin{figure*}
	\centering
	\includegraphics[width=16cm]{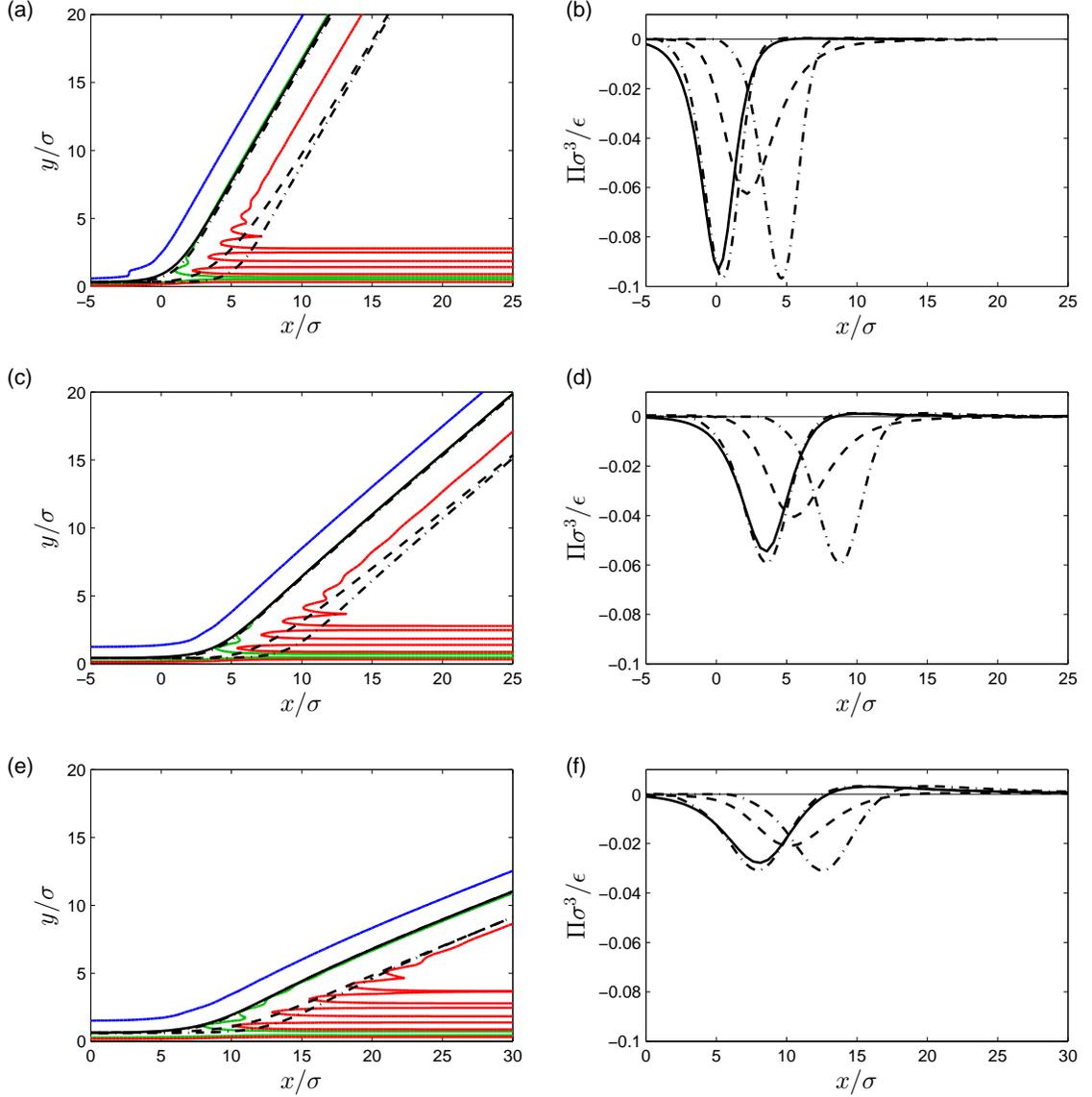}
	\caption{Density contours \new{(left column, figures (a),(c),(e))} and disjoining pressures \new{(right column, figures (b), (d), (f))} for three contact line
    regions. The top, middle and bottom rows depict results for the three different
    substrate strengths $\alpha_{\text{w}} \sigma^3/\depthLJ = \{1.25,1.375,1.47\}$, corresponding
    to Young contact angles $60.0^\circ$, $41.1^\circ$ and $20.4^\circ$,
    respectively. In the left column, the contour lines correspond to number densities $\klamm{\nDensity -
    \nDensityV}/\klamm{\nDensityL - \nDensityV} = \{0.05,0.5,0.95\}$ from left to right, respectively.	Height profiles
    $h_{\text{II}}$ and $h_{\text{III}}$ as in (\ref{eq:2DDisjoiningPressureIntegrated_hP}) and
    (\ref{eq:Def_HIII}) are represented by the dashed and solid lines,
    respectively. $h_{\text{I}}$ as defined by
    (\ref{eq:2DDisjoiningPressureIntegrated_hP:H}) is plotted with dash-dotted
    lines twice, to match $h_{\text{II}}$ and $h_{\text{III}}$ for large film
    thicknesses, through a corresponding choice of $h_B$ in (\ref{eq:BoundaryConditionHI}). \new{We note
    that in the left column,} the solid, one of the dash-dotted lines and the density isoline representing
		$\klamm{\nDensity - \nDensityV}/\klamm{\nDensityL - \nDensityV} = 0.5$ are virtually
    indistinguishable. \new{The right column depicts} the disjoining pressure profiles. The
    dashed line represents $\DisjoiningPressure_{\text{II}}\klamm{x}$ as defined in
    Eq.~(\ref{eq:SumRuleDisjoiningPressue2D}), the dash-dotted line represents
    $\DisjoiningPressure_{\text{I}}\klamm{h_{\text{I}}(x)}$ for the two shifted cases
    $h_{\text{I}}$ as depicted in the \new{left column}, and employing data of the
    adsorption isotherm for the disjoining pressure, using
    (\ref{eq:DisjoiningPressureAdsorptionIsotherm}). For comparison with
    Eq.~(\ref{eq:FilmHeightAndDisjoiningPressure}), the scaled curvature
    $-\surfaceTensionLV \dI\klamm{h_{\text{III}}'/\sqrt{1+(h_{\text{III}}')^2} }/\dI
    x$ is plotted with the solid line.}
    \label{fig:2014_1_7_17_31_DisjoiningPressureX}
\end{figure*}

In Fig.~\ref{fig:2014_1_7_17_31_DisjoiningPressureX}, we show results of the
equilibrium DFT-FMT computations in the contact line region for three
different substrate strengths, together with plots of the height profiles
$h_{\text{I}}$, $h_{\text{II}}$ and $h_{\text{III}}$ and the disjoining
pressure profiles $\DisjoiningPressure_{\text{I}}$ and
$\DisjoiningPressure_{\text{II}}$. We note that we have plotted
$h_{\text{I}}$ twice, to match $h_{\text{II}}$ and $h_{\text{III}}$ for large
film thicknesses, by use of the corresponding boundary condition~(\ref{eq:BoundaryConditionHI}). It is also worth noting that through
Eq.~(\ref{eq:FilmHeightAndDisjoiningPressure}), the disjoining pressures
correspond to the \new{scaled curvatures} of the height profiles. In all
cases, the numerical results show an excellent agreement of
\new{sum rules
(\ref{eq:FrumkinResultDisjoiningPressure1}) and (\ref{eq:SumRuleNormalForceBalance}).
This is shown in Table
\ref{tab:SumRuleResults}, where we compare the contact angles obtained by
evaluating  sum rules (\ref{eq:FrumkinResultDisjoiningPressure1}) and
(\ref{eq:SumRuleNormalForceBalance}) through the limiting behavior $\lim_{x \to \infty} h_{\text{I/II}}(x)$, with the Young contact angle
$\theta_{\text{Y}}$. For ease of comparison, let us define}
\begin{align}
\new{\theta_{\text{Y,I/II}} = \lim_{x \to \infty} \tan^{-1} \klamm{ h'_{\text{I/II}}(x) } .} \label{eq:DefineLimitingAngles}
\end{align}
 \new{We note that, as the height profiles $h_{\text{I,II}}$ are defined through Eq.~(\ref{eq:FilmHeightAndDisjoiningPressure}), the sum rules (\ref{eq:SumRuleNormalForceBalance}) and (\ref{eq:FrumkinResultDisjoiningPressure1})  lead to the same limiting contact angles for each of the height profiles.}

\begin{table}[htdp]
\begin{center}
\begin{tabular}{c|c|c|c}
$\alpha_{\text{w}}\sigma^3/\varepsilon$ & $\theta_{\text{Y}}$ & $\new{\theta_{\text{Y,I}}}$ & $\new{\theta_{\text{Y,II}}}$\\\hline
$1.25$ & $60.0^\circ$ & $60.0^\circ$ & $59.1^\circ \pm 1.7^\circ$ \\
$1.375$ & $41.1^\circ$ &$41.1^\circ$ & $39.9^\circ \pm 2^\circ$ \\
$1.47$ & $20.4^\circ$ & $20.4^\circ$ & $22.1^\circ \pm 2^\circ$
\end{tabular}
\end{center}
\caption{The Young contact angle $\theta_{\text{Y}}$ in
Eq.~(\ref{eq:YoungsEquation}) is compared
\new{with the contact angles obtained from the limiting behavior of the slope of the height profiles
$h'_{\text{I/II}}$, defined in Eq.~(\ref{eq:DefineLimitingAngles}) for
substrates of different strengths $\alpha_{\text{w}}$. As the height profiles
are defined through Eqs.~(\ref{eq:2DDisjoiningPressureIntegrated_hP:H}) and
(\ref{eq:2DDisjoiningPressureIntegrated_hP}), respectively, this amounts to
an error-check of sum rules (\ref{eq:FrumkinResultDisjoiningPressure1}) and
(\ref{eq:SumRuleNormalForceBalance}), respectively. Note that both height
profiles $h_{\text{I/II}}$ satisfy
Eq.~(\ref{eq:FilmHeightAndDisjoiningPressure}), which means that both sum
rules (\ref{eq:FrumkinResultDisjoiningPressure1}) and
(\ref{eq:SumRuleNormalForceBalance}) lead to the same limiting contact angle
for each of the height profiles. Error bounds for $\theta_{\text{Y,II}}$ were
estimated employing the numerical error in the computation of
$\DisjoiningPressure_{\text{II}}(\pm \infty)$.}} \label{tab:SumRuleResults}
\end{table}%

The density profiles in Figs.~\ref{fig:y2MaxAsymptotics60}
and~\ref{fig:2014_1_7_17_31_DisjoiningPressureX} reveal the structure of the
fluid in the immediate vicinity of the contact line. It is evident that the
fluid particles are densely packed close to the wall at the wall-liquid
interface due to hard-sphere effects. In particular, the transition between
the wall-vapor interface and the wall-liquid interface seems to lead to a
quasi step-like increase of the density. This influences the structure of the
liquid-vapor interface in the vicinity of the contact line. As attraction
with the wall increases and the contact angle decreases, packing close to the
wall becomes even more pronounced. Most importantly, we observe how the
structure of the liquid-vapor interface is significantly perturbed close to
the wall due to hard-sphere packing effects and ultimately merges with the wall-vapor interface ahead of the macroscopic liquid wedge.
Finally, we see that for $y
> 5\sigma$, the film height based on the adsorption, $h_{\text{III}}$, seems
to coincide for all three cases with the isodensity line for
$\klamm{\nDensity - \nDensityV}/\klamm{\nDensityL - \nDensityV} = 0.5$.

Let us now look at the results for the film heights $h_{\text{I}}$ and
$h_{\text{II}}$. We can make two main observations. First, there seems to be
a reasonable agreement between height profiles $h_{\text{I}}$ and
$h_{\text{III}}$ \new{in Figs.~\ref{fig:2014_1_7_17_31_DisjoiningPressureX}
(a), (c) and (e).}
A more accurate means to compare the behavior of the height profiles is through their corresponding disjoining pressure profiles in
Figs.~\ref{fig:2014_1_7_17_31_DisjoiningPressureX} (b), (d) and (f). Note
that the corresponding disjoining pressure plots correspond to the rescaled
curvatures of the height profiles, in accordance with
Eq.~(\ref{eq:FilmHeightAndDisjoiningPressure}). We observe that the maximal
curvature of both height profiles $h_{\text{I}}$ and $h_{\text{III}}$ agree
very well. Also, the curvature of both height profiles $h_{\text{I}}$ and
$h_{\text{III}}$ changes sign, which is more evident in
Fig.~\ref{fig:2014_1_7_17_31_DisjoiningPressureX} (f). In contrast,
$h_{\text{II}}$ exhibits a lower curvature than $h_{\text{I}}$ and
$h_{\text{III}}$ and it does not change its sign.
Furthermore, $h_{\text{II}}$ approaches an isodensity line for
$\klamm{\nDensity - \nDensityV}/\klamm{\nDensityL - \nDensityV}$ around
$0.95$, i.e.~much greater than $0.5$. We note that the results for
$h_{\text{II}}$ are similar to results obtained in
Ref.~\onlinecite{Herring:2010vn} for fluids with short-ranged interactions
and using MC computations in that the height profile $h_{\text{II}}$
approaches isodensity lines $\klamm{\nDensity - \nDensityV}/\klamm{\nDensityL
- \nDensityV} \approx 0.95$ for large $x$.

At the same time the results are surprising for two reasons. First, the
height profile $h_{\text{I}}$ is defined through the disjoining pressure
$\DisjoiningPressure_{\text{I}}$, which is based on computations of planar
wall-fluid interfaces. Hence, it loses some of the physics associated with
the true 2D contact line region profiles in that it does not include any
nonlocal effects in the direction parallel to the substrate, or effects due
to the slope of the liquid-vapor interface. Nevertheless, it does give a
good prediction of the adsorption height profile $h_{\text{III}}$ for contact
angles up to $60^\circ$. Second, the height profile $h_{\text{II}}$, which is
based on the disjoining pressure $\DisjoiningPressure_{\text{II}}$, seems to
behave very differently in the vicinity of the contact line compared to the
DFT-FMT computations, even though it contains information from the full 2D
density distribution.

The computations of height profiles through a three-phase contact line bring
us a considerable step closer to addressing one of the main questions posed
in the discussion papers that appeared in~{\it Eur. Phys. J.: Spec. Top.},
special issue `Wetting and Spreading Science - quo vadis?',~\cite{macdowell2011computer,Henderson:2011:EPJST:DisjoiningPressure,MacDowell:2011:ResponseEPJST,Henderson:EPJST:ResponseMacDowell,Henderson:NoteContinuingContactLine}
which is: Considering the disjoining pressure based on the normal force
balance $\DisjoiningPressure_{\text{II}}(x)$, which is the correct choice of
order parameter $\filmThickness$, if indeed there is one, that gives an
accurate local function $\DisjoiningPressure_{\text{II}}(\filmThickness)$?
In this special issue, Henderson~\cite{Herring:2010vn,Henderson:EPJST:ResponseMacDowell} noted that the disjoining pressure is inherently
non-local, and that there is no unique pair
$\klamm{\DisjoiningPressure,\filmThickness}$, in accordance with
Parry~\textit{et~al.}~\cite{parry2006derivation} In this context,
MacDowell~\cite{MacDowell:2011:ResponseEPJST} notes that the nonlocality of
the disjoining pressure can only matter very close to the contact line, as
far enough from the contact line Young's equation must be satisfied.

The computations presented in this study are a decisive first step towards addressing the question of
universality of the disjoining pressures $\DisjoiningPressure_{\text{I}}$ and
$\DisjoiningPressure_{\text{II}}$.
\new{In particular, we show that for the cases considered here, the disjoining pressure obtained from
the adsorption isotherm seems to accurately predict the height profile even
very close to the contact line. However, we also note that the disjoining
pressure based on the adsorption isotherm apparently does not correspond to
the excess pressure acting on the substrate obtained from a normal force
balance. One way to test if there is a unique pair
$\klamm{\DisjoiningPressure,\filmThickness}$ would be by comparing} static nanodroplets with each of the two
disjoining pressures and compare the results to those obtained from DFT-FMT,
but this is beyond the scope of the present study.

\section{Conclusion}

We have computed the density distribution in the vicinity of a three-phase
contact line at equilibrium using a DFT-FMT theory with a mean-field
Barker-Henderson approach for long-range particle interactions for three
different substrate strengths, corresponding to contact angles of
$20.4^\circ$, $41.1^\circ$ and $60.0^\circ$. We have confirmed that the
results satisfy basic sum rules to a good accuracy.
The computations allow us to probe the fluid structure in the immediate
vicinity of the contact line: Fluid particles are closely packed close to the
contact line due to hard-sphere effects. This packing leads to a
quasi-stepwise increase of the density as the wall is approached. For smaller
contact angles, i.e. as the attraction of the wall increases, the stepwise
structure of the density is amplified.

Furthermore, we have employed numerical results of adsorption isotherms for
different substrate strengths to define a disjoining pressure in the spirit
of Derjaguin-Frumkin adsorption theory. We have also used the results from
our equilibrium DFT computations of the density profile to define a
disjoining pressure based on the normal force balance at the contact line.
Via an effective mean-field Hamiltonian approach, both disjoining pressures
were employed to define height profiles to describe the three-phase contact
line. These were compared with the height profile defined by the adsorption
of the equilibrium density obtained from DFT.

The results of the comparison of the two disjoining pressures can be
summarized as follows: The disjoining pressure based on the adsorption
isotherm following the Derjaguin-Frumkin theory shows good agreement with the
DFT adsorption height profile in terms of maximal curvature and behavior for
large film heights. In contrast, the height profile defined through the
disjoining pressure based on a normal-force balance shows a very different
behavior, in particular, its maximal curvature is lower than that obtained
from DFT and Derjaguin-Frumkin, and it shows a different behavior for large
film heights compared to the other height profiles.
\new{Our results hence show that the disjoining pressure definition which gives the
better prediction for the adsorption film thickness is based on the
adsorption isotherm and does not correspond to the excess pressure acting on
the substrate, thus contradicting the classical notion of what the disjoining
pressure stands for.}

\new{One important restriction of the model used in this work is that it is of a mean-field type which does not
take into consideration thermal fluctuations.~\cite{ArcherEvans:2013,EvansHendersonHoyle:1993,MacDowellBenet:2014}
While we do not expect this 
to alter the general results of this work, fluctuations cannot be neglected
generally. For example, fluctuations were observed when modeling a contact
line using an MC algorithm for a fluid with short-range fluid-fluid
interactions.~\cite{Herring:2010vn} We note that including fluctuations in the fluid
description calls for an amended Hamiltonian theory, in which the
liquid-vapor interface has to be assumed to depend on the film thickness,
such as suggested by MacDowell~\textit{et~al.} for long-range fluid-fluid
interactions.~\cite{MacDowellBenet:2014} However, incorporating fluctuations
in a DFT model is highly nontrivial~\cite{ArcherEvans:2013} and beyond the
scope of this work.}

Clearly, there are many future directions that can be explored. For
instance, how chemically and/or topographically heterogeneous substrates,
which are known to influence wetting characteristics
substantially,~\cite{Nikos2009,Nikos2010,Nikos2011a,Nikos2011b,Raj2011} affect
the fluid structure in the vicinity of the contact line. Of particular
interest would also be the much more involved dynamic case. For this purpose,
the dynamic DFT approach developed
recently for colloidal fluids~\cite{Ben2012a,Ben2012b,Ben2013a,Ben2013b} should serve as a basis
for the accurate modeling of moving contact lines as it
takes into account both microscale inertia and hydrodynamic interactions, two effects which strongly influence nonequilibrium properties.
We shall address these and related issues in future studies.

\section{Acknowledgments}
We acknowledge financial support from ERC Advanced Grant No. 247031 and Imperial College through a DTG International Studentship.

\end{document}